\documentclass[aps,prl,twocolumn,groupedaddress,amssymb]{revtex4-2}
\makeatletter
\def\set@curr@file#1{%
  \begingroup
    \escapechar\m@ne
    \xdef\@curr@file{\expandafter\string\csname #1\endcsname}%
  \endgroup
}
\def\quote@name#1{"\quote@@name#1\@gobble""}
\def\quote@@name#1"{#1\quote@@name}
\def\unquote@name#1{\quote@@name#1\@gobble"}
\makeatother
\usepackage{graphicx}% Include figure files
\usepackage{dcolumn}% Align table columns on decimal point
\usepackage{bm}% bold math
\usepackage{amsmath}%Align equations
\usepackage{xcolor}
\usepackage{epstopdf}
\usepackage{hyperref}
\usepackage{mwe} % new package from Martin scharrer
\usepackage{ulem}

\usepackage[utf8]{inputenc}

\begin{document}

\title{Simple crowd dynamics to generate complex temporal contact networks}
\author{Razieh Masoumi\textsuperscript{1,2}}
%\email{rmrmasoomi@gmail.com}
\author{Juliette Gambaudo\textsuperscript{1}}
\author{Mathieu Génois\textsuperscript{1}}
\email{mathieu.genois@cpt.univ-mrs.fr}

\affiliation{\textsuperscript{1} Aix-Marseille Universit\'e, CNRS, CPT, Marseille, France}
\affiliation{\textsuperscript{2} Department of Molecular Medicine, University of Padova, Italy.}

\date{\today}

\begin{abstract}
  Empirical contact networks exhibit peculiar characteristics stemming from social, psychological, physical mechanisms governing human interactions. In this study, we test whether we are able to reproduce some of their dynamical properties from relatively simple models of 2D particle dynamics (random walk, active brownian particles and Vicsek model). Among the temporal properties of contact networks, we show that the distribution of inter-contact durations can be recovered with these simple models, and that this property is simply related to the well-know first-return process, which explains the $-3/2$ exponent that is found in both the numerical models and empirical contact networks.
%Empirical contact networks or interaction networks demonstrate peculiar characteristics stemming from the fundamental social, psychological, physical mechanisms governing human interactions. Although these mechanisms are complex, we test whether we are able to reproduce some dynamical properties of these empirical networks from relatively simple models. In this study, we perform simulations for a range of 2D models of particle dynamics, namely the Random Walk, Active Brownian Particles, and Vicsek models, to generate artificial contact networks. We investigate temporal properties of these contact networks: the distributions of contact durations, inter-contact durations and number of contact per pair of particle. We demonstrate that the distribution of inter-contact durations can be recovered by the dynamics of these simple crowd particle models, and show that it is simply related to the well-know first-return process, which explains the $-3/2$ exponent that is found in both the numerical models and empirical contact networks.
\end{abstract}

%\keywords{}

\maketitle

\section{Introduction}
Temporal networks, characterized by dynamic interactions and evolving connections over time, have gained significant attention across various disciplines in recent years, precisely because of their evolving nature. These networks capture essential aspects of real-world systems, where relationships, interactions, and information flows are highly time-dependent. Analyzing temporal networks provides valuable insights into numerous fields, including ecology\cite {bajardi2011dynamical, tantipathananandh2007framework, sundaresan2007network, croft2004social, blonder2011time}   disease spreading \cite{lee2012exploiting, volz2007susceptible, karsai2011small}, transportation \cite{pan2011path}, neuroscience \cite{bassett2013task, mantzaris2013dynamic, bassett2015learning}, communication networks such as mobile phone calls \cite{jiang2013calling, moody2005dynamic, kivela2012multiscale} and email networks \cite{eckmann2004entropy, ebel2002scale}, as well as citation networks \cite{rosvall2010mapping, rosvall2014memory}.

The structures and properties of temporal networks constrain processes which unfold on top of them. While understanding the effect of these structures is usually done with randomisation techniques \cite {holme2005network, holme2014analyzing, holme2012temporal, Mathieu2022}, generative models are needed to understand their origin.
In the realm of generative models for temporal networks, there exist several noteworthy approaches, including but not limited to the following approaches.
Holme introduced a fundamental technique known as ``Static networks with link dynamics'' \cite{holme2013epidemiologically, rocha2013bursts}. In this approach a temporal network is created by firstly generating a static network using a specific model and then generating a sequence of contacts for each link, typically without considering the network position of the links.
Perra et al. \cite{Perra_2012} introduced an even more simplified model of temporal networks using a graph sequence framework, called activity-driven networks \cite{didier}. Other approaches tried to take into account the intrinsic heterogeneous properties of the dynamics of empirical networks by implementing reinforcement processes, such as link-node memory models \cite{vestergaard2014memory}, and Self-exciting point processes \cite{masuda2013self, cho2013latent}.

Among temporal networks, proximity networks, which describe how individuals interact with each other in the physical space, are of particular interest for various contexts such as disease transmission, information dissemination, crowd management during disasters, and more. 
In this regard, the study of temporal networks from a socio-temporal perspective has acquired significant attention in recent years, primarily linked to the emergence of contemporary data collection techniques such as  GPS \cite{crofoot2011aggression}, and RFID chips \cite{barrat2013empirical, kibanov2014temporal, panisson2013fingerprinting,voirin2015combining}.  With the aid of advanced technology and data collection techniques, researchers can now analyze these networks in great detail, shedding light on the dynamics and patterns that govern our social exchanges. In this context, a groundbreaking investigation was the Reality Mining project, which involved outfitting Massachusetts Institute of Technology students with cell phones. These phones were equipped with Bluetooth technology capable of sensing their closeness to fellow individuals \cite{eagle2006reality}. Another relevant effort in this area is the SocioPatterns Project, which has devised a mechanism for gauging physical closeness through wearable badges incorporating radiofrequency identification devices (RFID) \cite{cattuto2010dynamics}. Within this initiative, they have successfully constructed closed gathering temporal networks for various groups, including hospital patients \cite{isella2011close}, conferences participants \cite{stehle2011simulation}, and students in schools \cite{ficarola2015capturing}.

As these networks arise from interactions between agents in the physical space, they may be intrinsically different form other classes of temporal networks which emerge from non spatial interactions (such as communication networks for example). Among generative models for temporal networks, few have tried to tackle the class of face-to-face interaction networks specifically, and thus a comprehensive theoretical understanding of face-to-face interaction data remains elusive.
However, the underlying spatial aspect of their origin allows to take a different approach in their generative process.
Starnini's research, which introduces a dynamic framework for human interactions, is a pioneering work in this field \cite{starnini2013modeling}. The framework utilizes a two-dimensional random walk, with each agent possessing an attractiveness factor that influences the pace of movement for individuals in their proximity. This approach provides valuable insights into the dynamics of human interactions. However, the relation between spatial constraints and the properties of the temporal network of interactions remain unclear.

Previous analysis of empirical contact networks, and in particular the SocioPatterns datasets, have shown that specific dynamic features such as contact duration and inter-contact duration distributions exhibit similar properties across various contexts \cite{starnini2013modeling}.
While acknowledging the inherent complexities of these networks, a compelling question arises: Can we capture some of their dynamic characteristics using simpler contact models?
In this regard, we explore the feasibility of understanding empirical contact networks through the lens of relatively simple contact models which take into account the spatial origin of the contacts. Specifically, we focus on 2D particle dynamics, and we investigate the behavior of three distinct models: the Random Walk, the Active Brownian Particle model, and the Vicsek model. The first one is the simplest model for the dynamics of particles; the second one allows to introduce some limited memory effect on the trajectory, and the third one is one of the simplest models for motion of active matter exhibiting collective behavior.

Our primary objective is to generate contact networks from each of these 2D particle models and analyze the temporal aspects of these networks. We focus on three distributions: contact duration, inter-contact duration, and the number of contacts per pair of particles. By scrutinizing these temporal patterns, we aim to discern which aspects of empirical contact networks can be elucidated through the dynamics of these straightforward crowd particle models and which aspects necessitate additional mechanisms or complexities to achieve accurate representation.

One key finding of our investigation is the consistent presence of heavy-tailed distributions in the inter-contact duration, irrespective of the specific particle model employed. This interesting observation suggests that, though sociological, psychological, neurological and biological mechanisms are at play in human interactions, some patterns can be explained by a very simple, purely statistical effect. Indeed, we manage to connect the shape of the distribution of inter-contact durations to a simple first-return time problem, and recover the empirical exponent of $-1.5$ observed in all models and datasets.

\section{\label{sec2}Methods}
\subsection{\label{subsec1}Models}
Our primary objective is to understand the properties of empirical contact networks by exploring what happens for basic interaction models. One question in particular we aim to address is whether there exist universal properties that remain independent while varying the contact dynamics. To achieve this, we use three distinctive two-dimensional particle dynamics models: Random Walk, Active Brownian Particle, and Vicsek model.
In all models we first assume that the particles are point-wise entities, which means their presence does not influence each other; their trajectories remain unchanged when they meet. In a second version, particles have a fixed size, and simply stop at contact.

We consider in all cases a system containing $N=1000$ point-wise particles confined within a 2D box of side length $L=100$. We have chosen this particular density for the particles to ensure comparability with the density of people in the conference data sets we have utilized \cite{Mathieu2022}. After setting random initial conditions, the particles are observed to move freely within the box. When they encounter the boundaries, they undergo classic reflection due to the reflective boundary conditions. For completeness, we further tested the models with periodic boundary conditions or on an infinite space.

We deliberately adopt these relatively simple particle models and make their contact network, aiming to uncover potentially reproducible properties of empirical contact networks, even within the context of basic pedestrian models. While these models are widely explored, we aim to provide a concise yet comprehensive explanation of each of them, providing a clear understanding of their characteristics and relevance to our study.

\subsection{\label{subsec2}Contact networks}
A social interaction can mean many different behaviors, such as conversation, physical or eye contact, all of which hold significance in analyzing connections within a crowd. In our scenario we focus on a simpler interpretation, namely what we call a ``contact'', which refers specifically to a physical closeness event. While physical proximity does not guarantee an interaction, past research indicates that it serves as a reliable indicator for an examination of the social context's structure \cite{schaible2021sensing}.

Such an approach was used by the SocioPatterns platform \cite{cattuto2010dynamics} for gathering empirical data about contacts between individuals. This platform has been extensively employed in the past decade to investigate interaction patterns in social settings \cite{vanhems2013estimating, kiti2016quantifying, genois2018can, kontro2020combining, ozella2021using, oliveira2022group}. The system involves sensors affixed to participants' nametags and antennas strategically positioned throughout the studied location to collect contact data from these sensors. Each sensor is equipped with an RFID chip which is able to detect other sensors within a proximity of approximately 1.5 meters. Notably, detection only occurs when two individuals are face-to-face, within each other's front half-spheres, as the emitted signal is blocked by the human body. A contact event is defined by such proximity and geometry. These contacts are recorded at 20-second intervals and are capped at 40 simultaneous contacts for an individual within a 20-second time frame. According to the system design, contacts lasting at least 20 seconds are certain to be recorded, while shorter contacts may also be recorded with a probability that diminishes as their duration decreases. 

While many different situations have been studied with this equipment, as a reference for the properties of empirical contact networks we utilized data collected with the SocioPatterns platform during 4 face-to-face interactions conferences \cite{Mathieu2022}. These data sets were preferred as they are characterised by an adult population, larger freedom of movement and less schedule constraints on the individuals' behavior.

In order to be able to compare empirical and artificial interactions, we detect contacts in the models in a way that mirrors the contact definition from the SocioPatterns data: two particles are in contact when they are closer than the detection radius and facing each other (see SI for details).

To compare empirical and artificial contact networks, we focus on the distributions of three temporal properties of the contacts, namely contact duration (number of consecutive time steps during which two particles are in contact), inter-contact duration (number of consecutive time steps during which two particles are not in contact) and the number of contacts per pair of particles.

\section{\label{sec4} Result }
\begin{figure*}[t]
	\centering
	\includegraphics[width=\textwidth]{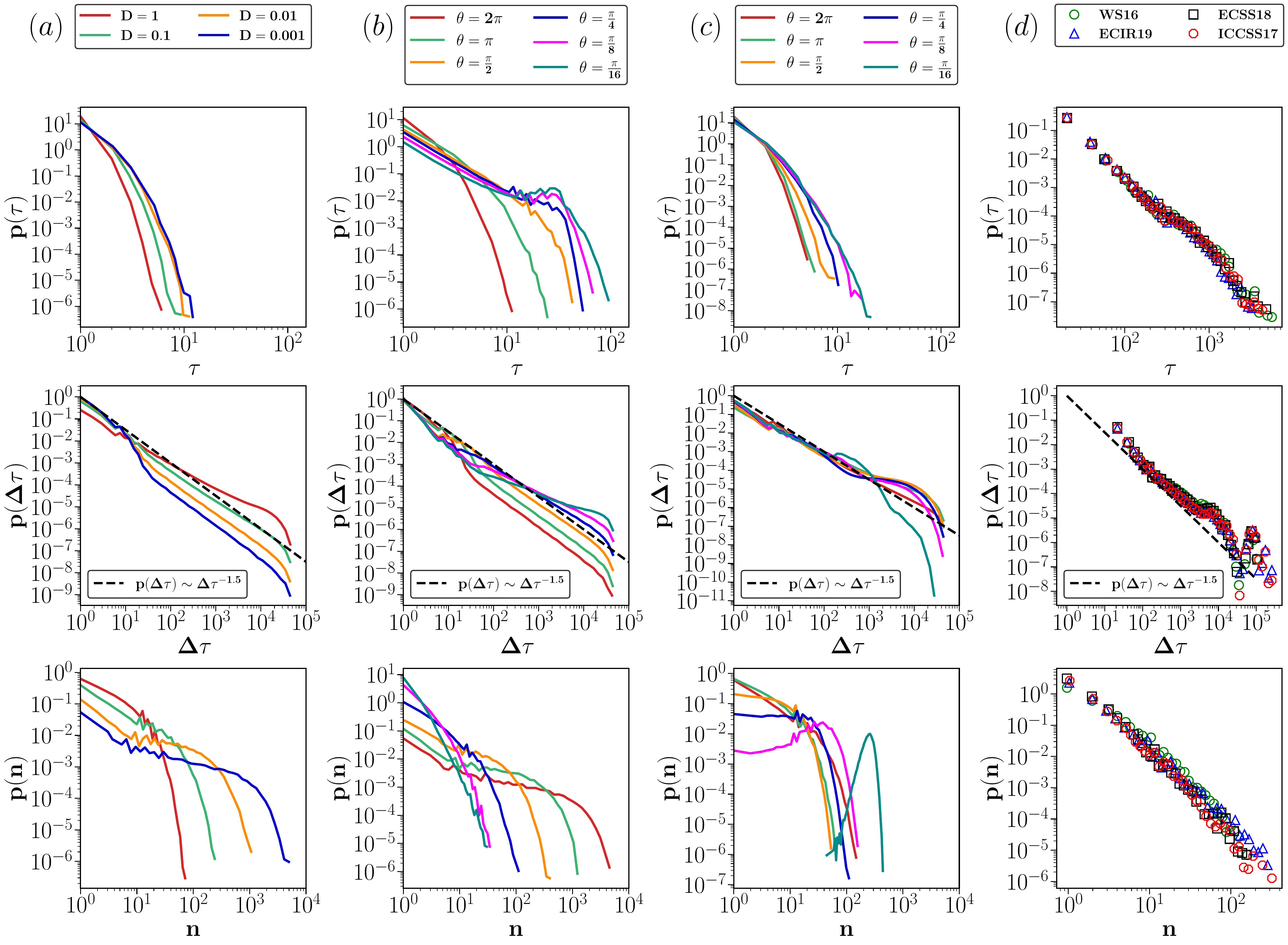} % Change this to your actual figure
	\caption{\textbf{Comparison of the distributions of contact duration, inter-contact duration, and number of contacts for the different models.} (a) 2D Random Walk, (b) Active Brownian Particles, (c) Vicsek model, (d) conference data.}
	\label{all-dist-final.eps}
\end{figure*}

\subsection{Temporal distributions for simple dynamics}

Fig. \ref{all-dist-final.eps} presents the contact duration, inter-contact duration, and number of contacts distributions for 2D Random Walk (2DRW), Active Brownian Particle (ABP) and Vicsek (VM) models respectively, for point particles and hard boundary conditions, and compares them with their empirical counterparts. In each model we tested several values for the parameter that fixes the level of noise (the diffusivity $D$ for 2DRW, the angle range $\theta$ for ABP and VM).

\subsubsection{Contact duration and number of contacts}

Figure \ref{all-dist-final.eps} (a) shows the distributions for the 2DRW for various diffusion coefficients: $D=1, 0.1, 0.01, 0.001$. As expected, the distributions have an exponential shape, directly due to the uncorrelated randomness of the process. When $D$ is large, particles easily escape the detection area of each other; consequently, they tend to stay in contact for shorter duration compared to when the diffusion coefficient is lower. This effect explains the relative positions of the distributions of contact duration. Similarly, distributions of number of contacts per pair of particles also exhibit an exponential tail and an ordering according to diffusivity. When two particles with a small diffusion coefficient break contact, they are less likely to move away quickly; as a result, they have a higher probability of making contact again in a shorter time period. Consequently, in this scenario, the number of contacts between two specific particles increases as the diffusivity decreases.

Figure \ref{all-dist-final.eps} (b) shows the distributions for the ABP model. In this model, except for the case $\theta = 2\pi$ which is similar to the 2DRW, when particles come into contact with each other they tend to remain in contact for a longer duration, especially as $\theta$ decreases. This is mostly due to the fact that, when $\theta$ is small, particles in contact may have aligned directions, which are maintained for a while, leading to longer-lasting interactions. Contrarily, the ballistic aspect of their trajectories ensures that when particles break contact, they move away and stay apart longer, leading to a decrease in the total number of contacts as $\theta$ decreases.

Figure \ref{all-dist-final.eps} (c) shows the distributions for the Vicsek model. This model is peculiar as it undergoes a phase transition as $\theta$ varies, from a disordered state for high values of $\theta$ to a flocking phase for low values of $\theta$, with a critical value of approximately $\frac{\pi}4$ in our case (see SI). In the flocking phase, particles exhibit longer contact duration, as they move in groups, have aligned directions and thus stay close to each other. As $\theta$ increases, the alignment effect vanishes and the contact durations decrease. In all cases though, the shape of the distribution remains exponential, similar to the 2DRW case. The behavior of the distribution of number of contacts is different, with a non-linearity directly due to the phase transition. While in the disordered phase the distribution is exponential, its shape changes towards higher values of $n$ in the flocking phase, due to the numerous contacts which occur in the flocks.

As seen by comparing the previous results with Figure \ref{all-dist-final.eps} (d), which shows the distributions for the four conferences, all of them are completely different from what is observed for empirical contacts. Empirical distributions exhibit a power law shape, which has been shown to be a characteristic commonly observed in human behavior. This discrepancy is not a surprise, as the models presented here incorporate dynamics that are far from being realistic. The shapes of the distributions are intricately tied to the specific dynamics inherent to each particle model, demonstrating the distinctive characteristics of each model and its interactions.

\subsubsection{Inter-contact duration}

However, we do observe something unexpected when examining the distributions of inter-contact duration. In all three models, this distribution exhibits the same power-law shape as it does in the data sets, and always with a similar exponent of $-1.5$ which is also observed for empirical contacts. For each model, this shape holds for all values of the parameter. The exponent is also the same in most cases, changing only slightly in the case of high diffusivity for 2DRW or long persistence for ABP. This remarkable fact seems to indicate that this particular property of contact dynamics between moving particles is largely independent from the details of the dynamics. Changing from point-particles to particles with a size which stop at contact, or modifying the boundary conditions to bi-periodic or infinite space does not modify this property (see SI).

Focusing on the random walk of point-particles on a boundless space, arguments allow to relate the analytical derivation of this distribution to the distribution of first-passage times in a 1D process describing the evolution of the distance between a pair of particles.
In the Brownian limit, this distance approximates a one-dimensional diffusive process over short and long timescales \cite{Rast_2022} for which the first-passage time distribution is characterized by a power-law tail of exponent -3/2 \cite{Redner_2001}.

The fact that we numerically get the same exponent for the tail of the distribution with all boudary conditions and particle with or without sizes seems to indicate that the effect of these modifications does not change the core of the derivation. Furthermore, the fact that we get the same exponent in the other models (ABP and Vicsek before the phase transition) lets us formulate a general hypothesis about the universality of this effect: if the motion of particles is ``random enough'' to be statistically equivalent to a random walk, then the distribution of inter-contact duration exhibits a power-law tail with exponent $-3/2$.

Finally, the fact that we observe the same feature in empirical data of contacts between individuals lets us believe that the motion of human beings, although far from being a random walk at the individual's scale and at short time, is also ``sufficiently random'' to exhibit this property when considering a crowd over a certain time window.

%%%%%%%%%%%%%%%%%%%%%%%%%%%%%%%%%%%%%%%%%%%%%%%%%
\section{\label{sec4} Conclusion }

In this study we focused on generating contact networks using three distinct two-dimensional particle dynamics models: 2D Random Walk, Active Brownian Particles, and Vicsek model. These contact networks serve as representations of pairwise interactions between particles in crowd models, offering insights into collective behavior and emergent properties of crowds. The primary objective was to compare the properties of empirical contact networks with those of networks generated by these models, and determine whether some of them could be explained by such simple mechanisms.

The analysis involved investigating the distributions of contact duration, inter-contact duration, and the number of contacts for the various particle models. We observed that the different models exhibited various characteristics related to their microscopic dynamics.
However, the distribution of inter-contact duration exhibited in all cases a shape similar to what is found in empirical network of contacts, with a power-law tail of exponent $-1.5$. Further simulations showed that this feature was retained even for a 2D random walk of point particles on a infinite space, which allowed us to relate it to a first-return time distribution in a 1D random walk, for which the classical results indeed gives a $-3/2$ exponent.

The presence of this tail in all variants of the 2D Random Walk indicates that this behaviour is not affected by either the fact that particles have a size, nor by the boundary conditions. Furthermore, the tail is retained in the other models too. This demonstrates that, while the dynamics of the particles might be very different, they are statistically equivalent to the 2D Random Walk with respect to the inter-contact durations. Finally, the presence of the same behaviour in empirical data seems to indicate that the trajectories of individuals, while far from being random, are nonetheless also statistically equivalent to random walks with respect to inter-contact durations.

In conclusion, this study provided insights into the dynamics of contact networks, shedding light on behavior in crowds. The findings contribute to the understanding of universal properties in pairwise interactions, even within the context of basic crowd models. 

\section*{Acknowledgements}
\subsection*{Author contributions}
M.G. proposed the core idea of the project. R.M., J.G. and M.G. contributed to the scientific discussions of the work. R.M. and J.G. wrote code, performed numerical simulations and analyzed numerical and empirical data. R.M., J.G. and M.G. wrote and reviewed the final manuscript.

\subsection*{Funding}
R.M. is supported by an Emergence@INP grant from the CNRS. J.G. and M.G. are partially supported by the Agence Nationale de la Recherche (ANR) project DATAREDUX (ANR-19-CE46-0008).

\subsection*{Competing interests}
The authors declare no competing interests.

%%%%%%%%%%%%%%%%%%%%%%%%%%%%%%%%%%%%%%%%%%%%%%%%%%

\bibliographystyle{plain}  % Use the Nature bibliography style
\bibliography{mybib}

\begin{thebibliography}{10}

\bibitem{bajardi2011dynamical}
Paolo Bajardi, Alain Barrat, Fabrizio Natale, Lara Savini, and Vittoria
  Colizza.
\newblock Dynamical patterns of cattle trade movements.
\newblock {\em PloS one}, 6(5):e19869, 2011.

\bibitem{barrat2013empirical}
Alain Barrat, Ciro Cattuto, Vittoria Colizza, Francesco Gesualdo, Lorenzo
  Isella, Elisabetta Pandolfi, J~F Pinton, Lucilla Rav{\`a}, Caterina Rizzo,
  Mariateresa Romano, et~al.
\newblock Empirical temporal networks of face-to-face human interactions.
\newblock {\em The European Physical Journal Special Topics}, 222:1295--1309,
  2013.

\bibitem{bassett2013task}
Danielle~S Bassett, Nicholas~F Wymbs, M~Puck Rombach, Mason~A Porter, Peter~J
  Mucha, and Scott~T Grafton.
\newblock Task-based core-periphery organization of human brain dynamics.
\newblock {\em PLoS computational biology}, 9(9):e1003171, 2013.

\bibitem{bassett2015learning}
Danielle~S Bassett, Muzhi Yang, Nicholas~F Wymbs, and Scott~T Grafton.
\newblock Learning-induced autonomy of sensorimotor systems.
\newblock {\em Nature neuroscience}, 18(5):744--751, 2015.

\bibitem{blonder2011time}
Benjamin Blonder and Anna Dornhaus.
\newblock Time-ordered networks reveal limitations to information flow in ant
  colonies.
\newblock {\em PloS one}, 6(5):e20298, 2011.

\bibitem{cattuto2010dynamics}
Ciro Cattuto, Wouter Van~den Broeck, Alain Barrat, Vittoria Colizza,
  Jean-Fran{\c{c}}ois Pinton, and Alessandro Vespignani.
\newblock Dynamics of person-to-person interactions from distributed rfid
  sensor networks.
\newblock {\em PloS one}, 5(7):e11596, 2010.

\bibitem{cho2013latent}
Yoon-Sik Cho, Aram Galstyan, P~Jeffrey Brantingham, and George Tita.
\newblock Latent self-exciting point process model for spatial-temporal
  networks.
\newblock {\em arXiv preprint arXiv:1302.2671}, 2013.

\bibitem{crofoot2011aggression}
Margaret~C Crofoot, Daniel~I Rubenstein, Arun~S Maiya, and Tanya~Y Berger-Wolf.
\newblock Aggression, grooming and group-level cooperation in white-faced
  capuchins (cebus capucinus): Insights from social networks.
\newblock {\em American Journal of Primatology}, 73(8):821--833, 2011.

\bibitem{croft2004social}
Darren~P Croft, Jens Krause, and Richard James.
\newblock Social networks in the guppy (poecilia reticulata).
\newblock {\em Proceedings of the Royal Society of London. Series B: Biological
  Sciences}, 271(suppl\_6):S516--S519, 2004.

\bibitem{eagle2006reality}
Nathan Eagle and Alex Pentland.
\newblock Reality mining: sensing complex social systems.
\newblock {\em Personal and ubiquitous computing}, 10:255--268, 2006.

\bibitem{ebel2002scale}
Holger Ebel, Lutz-Ingo Mielsch, and Stefan Bornholdt.
\newblock Scale-free topology of e-mail networks.
\newblock {\em Physical review E}, 66(3):035103, 2002.

\bibitem{eckmann2004entropy}
Jean-Pierre Eckmann, Elisha Moses, and Danilo Sergi.
\newblock Entropy of dialogues creates coherent structures in e-mail traffic.
\newblock {\em Proceedings of the National Academy of Sciences},
  101(40):14333--14337, 2004.

\bibitem{ficarola2015capturing}
Francesco Ficarola and Andrea Vitaletti.
\newblock Capturing interactions in face-to-face social networks.
\newblock In {\em WEBIST}, pages 613--620, 2015.

\bibitem{Mathieu2022}
{Gauvin, Laetitia and G\'{e}nois, Mathieu and Karsai, M\'{a}rton and
  Kivel\"{a}, Mikko and Takaguchi, Taro and Valdano, Eugenio and Vestergaard,
  Christian L.}
\newblock Randomized reference models for temporal networks.
\newblock {\em SIAM Review}, 64(4):763--830, 2022.

\bibitem{genois2018can}
Mathieu G{\'e}nois and Alain Barrat.
\newblock Can co-location be used as a proxy for face-to-face contacts?
\newblock {\em EPJ Data Science}, 7(1):1--18, 2018.

\bibitem{holme2005network}
Petter Holme.
\newblock Network reachability of real-world contact sequences.
\newblock {\em Physical Review E}, 71(4):046119, 2005.

\bibitem{holme2013epidemiologically}
Petter Holme.
\newblock Epidemiologically optimal static networks from temporal network data.
\newblock {\em PLoS computational biology}, 9(7):e1003142, 2013.

\bibitem{holme2014analyzing}
Petter Holme.
\newblock Analyzing temporal networks in social media.
\newblock {\em Proceedings of the IEEE}, 102(12):1922--1933, 2014.

\bibitem{holme2012temporal}
Petter Holme and Jari Saram{\"a}ki.
\newblock Temporal networks.
\newblock {\em Physics reports}, 519(3):97--125, 2012.

\bibitem{isella2011close}
Lorenzo Isella, Mariateresa Romano, Alain Barrat, Ciro Cattuto, Vittoria
  Colizza, Wouter Van~den Broeck, Francesco Gesualdo, Elisabetta Pandolfi,
  Lucilla Rav{\`a}, Caterina Rizzo, et~al.
\newblock Close encounters in a pediatric ward: measuring face-to-face
  proximity and mixing patterns with wearable sensors.
\newblock {\em PloS one}, 6(2):e17144, 2011.

\bibitem{jiang2013calling}
Zhi-Qiang Jiang, Wen-Jie Xie, Ming-Xia Li, Boris Podobnik, Wei-Xing Zhou, and
  H~Eugene Stanley.
\newblock Calling patterns in human communication dynamics.
\newblock {\em Proceedings of the National Academy of Sciences},
  110(5):1600--1605, 2013.

\bibitem{karsai2011small}
M{\'a}rton Karsai, Mikko Kivel{\"a}, Raj~Kumar Pan, Kimmo Kaski, J{\'a}nos
  Kert{\'e}sz, A-L Barab{\'a}si, and Jari Saram{\"a}ki.
\newblock Small but slow world: How network topology and burstiness slow down
  spreading.
\newblock {\em Physical Review E}, 83(2):025102, 2011.

\bibitem{kibanov2014temporal}
Mark Kibanov, Martin Atzmueller, Christoph Scholz, and Gerd Stumme.
\newblock Temporal evolution of contacts and communities in networks of
  face-to-face human interactions.
\newblock {\em Science China Information Sciences}, 57:1--17, 2014.

\bibitem{kiti2016quantifying}
Moses~C Kiti, Michele Tizzoni, Timothy~M Kinyanjui, Dorothy~C Koech, Patrick~K
  Munywoki, Milosch Meriac, Luca Cappa, Andr{\'e} Panisson, Alain Barrat, Ciro
  Cattuto, et~al.
\newblock Quantifying social contacts in a household setting of rural kenya
  using wearable proximity sensors.
\newblock {\em EPJ data science}, 5:1--21, 2016.

\bibitem{kivela2012multiscale}
Mikko Kivel{\"a}, Raj~Kumar Pan, Kimmo Kaski, J{\'a}nos Kert{\'e}sz, Jari
  Saram{\"a}ki, and M{\'a}rton Karsai.
\newblock Multiscale analysis of spreading in a large communication network.
\newblock {\em Journal of Statistical Mechanics: Theory and Experiment},
  2012(03):P03005, 2012.

\bibitem{kontro2020combining}
Inkeri Kontro and Mathieu G{\'e}nois.
\newblock Combining surveys and sensors to explore student behaviour.
\newblock {\em Education Sciences}, 10(3):68, 2020.

\bibitem{didier}
Didier Le~Bail, Mathieu G\'enois, and Alain Barrat.
\newblock Modeling framework unifying contact and social networks.
\newblock {\em Phys. Rev. E}, 107:024301, Feb 2023.

\bibitem{lee2012exploiting}
Sungmin Lee, Luis~EC Rocha, Fredrik Liljeros, and Petter Holme.
\newblock Exploiting temporal network structures of human interaction to
  effectively immunize populations.
\newblock {\em PloS one}, 7(5):e36439, 2012.

\bibitem{mantzaris2013dynamic}
Alexander~V Mantzaris, Danielle~S Bassett, Nicholas~F Wymbs, Ernesto Estrada,
  Mason~A Porter, Peter~J Mucha, Scott~T Grafton, and Desmond~J Higham.
\newblock Dynamic network centrality summarizes learning in the human brain.
\newblock {\em Journal of Complex Networks}, 1(1):83--92, 2013.

\bibitem{masuda2013self}
Naoki Masuda, Taro Takaguchi, Nobuo Sato, and Kazuo Yano.
\newblock Self-exciting point process modeling of conversation event sequences.
\newblock {\em Temporal networks}, pages 245--264, 2013.

\bibitem{moody2005dynamic}
James Moody, Daniel McFarland, and Skye Bender-deMoll.
\newblock Dynamic network visualization.
\newblock {\em American journal of sociology}, 110(4):1206--1241, 2005.

\bibitem{oliveira2022group}
Marcos Oliveira, Fariba Karimi, Maria Zens, Johann Schaible, Mathieu
  G{\'e}nois, and Markus Strohmaier.
\newblock Group mixing drives inequality in face-to-face gatherings.
\newblock {\em Communications Physics}, 5(1):127, 2022.

\bibitem{ozella2021using}
Laura Ozella, Daniela Paolotti, Guilherme Lichand, Jorge~P Rodr{\'\i}guez,
  Simon Haenni, John Phuka, Onicio~B Leal-Neto, and Ciro Cattuto.
\newblock Using wearable proximity sensors to characterize social contact
  patterns in a village of rural malawi.
\newblock {\em EPJ Data Science}, 10(1):46, 2021.

\bibitem{pan2011path}
Raj~Kumar Pan and Jari Saram{\"a}ki.
\newblock Path lengths, correlations, and centrality in temporal networks.
\newblock {\em Physical Review E}, 84(1):016105, 2011.

\bibitem{panisson2013fingerprinting}
Andr{\'e} Panisson, Laetitia Gauvin, Alain Barrat, and Ciro Cattuto.
\newblock Fingerprinting temporal networks of close-range human proximity.
\newblock In {\em 2013 IEEE international conference on pervasive computing and
  communications workshops (PERCOM workshops)}, pages 261--266. IEEE, 2013.

\bibitem{Perra_2012}
N.~Perra, B.~Gon{\c{c}}alves, R.~Pastor-Satorras, and A.~Vespignani.
\newblock Activity driven modeling of time varying networks.
\newblock {\em Scientific Reports}, 2(1), jun 2012.

\bibitem{Rast_2022}
Mark~Peter Rast.
\newblock Contact statistics in populations of noninteracting random walkers in
  two dimensions.
\newblock {\em Phys. Rev. E}, 105:014103, Jan 2022.

\bibitem{Redner_2001}
Sidney Redner.
\newblock {\em A Guide to First-Passage Processes}.
\newblock Cambridge University Press, 2001.

\bibitem{rocha2013bursts}
Luis~EC Rocha and Vincent~D Blondel.
\newblock Bursts of vertex activation and epidemics in evolving networks.
\newblock {\em PLoS computational biology}, 9(3):e1002974, 2013.

\bibitem{rosvall2010mapping}
Martin Rosvall and Carl~T Bergstrom.
\newblock Mapping change in large networks.
\newblock {\em PloS one}, 5(1):e8694, 2010.

\bibitem{rosvall2014memory}
Martin Rosvall, Alcides~V Esquivel, Andrea Lancichinetti, Jevin~D West, and
  Renaud Lambiotte.
\newblock Memory in network flows and its effects on spreading dynamics and
  community detection.
\newblock {\em Nature communications}, 5(1):4630, 2014.

\bibitem{schaible2021sensing}
Johann Schaible, Marcos Oliveira, Maria Zens, and Mathieu G{\'e}nois.
\newblock Sensing close-range proximity for studying face-to-face interaction.
\newblock In {\em Handbook of Computational Social Science, Volume 1}, pages
  219--239. Routledge, 2021.

\bibitem{starnini2013modeling}
Michele Starnini, Andrea Baronchelli, and Romualdo Pastor-Satorras.
\newblock Modeling human dynamics of face-to-face interaction networks.
\newblock {\em Physical review letters}, 110(16):168701, 2013.

\bibitem{stehle2011simulation}
Juliette Stehl{\'e}, Nicolas Voirin, Alain Barrat, Ciro Cattuto, Vittoria
  Colizza, Lorenzo Isella, Corinne R{\'e}gis, Jean-Fran{\c{c}}ois Pinton,
  Nagham Khanafer, Wouter Van~den Broeck, et~al.
\newblock Simulation of an seir infectious disease model on the dynamic contact
  network of conference attendees.
\newblock {\em BMC medicine}, 9(1):1--15, 2011.

\bibitem{sundaresan2007network}
Siva~R Sundaresan, Ilya~R Fischhoff, Jonathan Dushoff, and Daniel~I Rubenstein.
\newblock Network metrics reveal differences in social organization between two
  fission--fusion species, grevy’s zebra and onager.
\newblock {\em Oecologia}, 151:140--149, 2007.

\bibitem{tantipathananandh2007framework}
Chayant Tantipathananandh, Tanya Berger-Wolf, and David Kempe.
\newblock A framework for community identification in dynamic social networks.
\newblock In {\em Proceedings of the 13th ACM SIGKDD international conference
  on Knowledge discovery and data mining}, pages 717--726, 2007.

\bibitem{vanhems2013estimating}
Philippe Vanhems, Alain Barrat, Ciro Cattuto, Jean-Fran{\c{c}}ois Pinton,
  Nagham Khanafer, Corinne R{\'e}gis, Byeul-a Kim, Brigitte Comte, and Nicolas
  Voirin.
\newblock Estimating potential infection transmission routes in hospital wards
  using wearable proximity sensors.
\newblock {\em PloS one}, 8(9):e73970, 2013.

\bibitem{vestergaard2014memory}
Christian~L Vestergaard, Mathieu G{\'e}nois, and Alain Barrat.
\newblock How memory generates heterogeneous dynamics in temporal networks.
\newblock {\em Physical Review E}, 90(4):042805, 2014.

\bibitem{voirin2015combining}
Nicolas Voirin, C{\'e}cile Payet, Alain Barrat, Ciro Cattuto, Nagham Khanafer,
  Corinne R{\'e}gis, Byeul-a Kim, Brigitte Comte, Jean-S{\'e}bastien Casalegno,
  Bruno Lina, et~al.
\newblock Combining high-resolution contact data with virological data to
  investigate influenza transmission in a tertiary care hospital.
\newblock {\em Infection Control \& Hospital Epidemiology}, 36(3):254--260,
  2015.

\bibitem{volz2007susceptible}
Erik Volz and Lauren~Ancel Meyers.
\newblock Susceptible--infected--recovered epidemics in dynamic contact
  networks.
\newblock {\em Proceedings of the Royal Society B: Biological Sciences},
  274(1628):2925--2934, 2007.

\end{thebibliography}

\end{document}